\documentclass[10pt,a4paper]{article}
\usepackage[utf8]{inputenc}
\usepackage[T1]{fontenc}
\usepackage{authblk}
\usepackage[]{graphicx}
\usepackage{hyperref}
\usepackage{mathtools}
\usepackage{booktabs}
\usepackage{amssymb}
\usepackage{color}
\usepackage{xspace}

\hypersetup{
    colorlinks=true,
    linkcolor=blue,
    citecolor=blue,
    urlcolor=blue,
}

\usepackage[backend=biber, citestyle=numeric-comp, sorting=none]{biblatex}
\AtBeginBibliography{\setlength{\emergencystretch}{2em}}

\newcommand{\mean}[1]{\ensuremath{\langle #1 \rangle}}
\newcommand{\WcmTwo}{\ensuremath{\mathrm{W\,cm}^{-2}}\xspace}
\newcommand{\eV}{\ensuremath{\mathrm{eV}}\xspace}
\newcommand{\HeIon}{\ensuremath{\mathrm{He}^{+}}\xspace}
\newcommand{\HeDIon}{\ensuremath{\mathrm{He}^{2+}}\xspace}
\newcommand{\HeTwoIon}{\ensuremath{\mathrm{He}_{2}^{+}}\xspace}
\newcommand{\Ee}{\ensuremath{E_{\mathrm{e}}^{\mathrm{max}}}\xspace}
\newcommand{\Ei}{\ensuremath{\mean{E_{\mathrm{i}}}}\xspace}
\newcommand{\Ne}{\ensuremath{N_{\mathrm{e}}}\xspace}
\newcommand{\Ni}{\ensuremath{N_{\mathrm{i}}}\xspace}

\title{
Correlated detection of electrons and ions emitted from laser-induced nanoplasmas
in doped helium nanodroplets in the single-shot single-cluster regime
}

\author[1,2]{Keshav Sishodia}
\author[3,4]{Cristian Medina}
\author[4]{Robert Moshammer}
\author[4]{Thomas Pfeifer}
\author[5,6]{Andreas Heidenreich}
\author[7]{Marcel Mudrich}
\author[1, *]{Sivarama Krishnan}

\affil[1]{Quantum Center of Excellence for Diamond and Emergent Materials, School of Interdisciplinary Sciences and Department of Physics, Indian Institute of Technology Madras, 600036 Chennai, India.}
\affil[2]{ELI Beamlines facility, The Extreme Light Infrastructure ERIC, Za Radnic\'{\i} 835, 253 41 Doln\'{\i} B\v{r}e\v{z}any, Czech Republic.}
\affil[3]{Department of Chemistry, University of California, Berkeley, CA, USA.}
\affil[4]{Max-Planck-Institut f{\"u}r Kernphysik, Saupfercheckweg 1, 69117 Heidelberg, Germany.}
\affil[5]{Kimika Fakultatea, Euskal Herriko Unibertsitatea (UPV/EHU) and Donostia International Physics Center (DIPC), P.K. 1072, 20080 Donostia, Spain}
\affil[6]{IKERBASQUE, Basque Foundation for Science, 48011 Bilbao, Spain}
\affil[7]{Institute of Physics, University of Kassel, 34132 Kassel, Germany.}
\affil[*]{srkrishnan@iitm.ac.in}

\date{}

\begin{document}
\maketitle

\begin{abstract}
We present single-cluster single-shot correlated measurements of electron velocity-map images and
ion time-of-flight traces from individual laser-induced nanoplasmas in xenon-doped helium nanodroplets.
Simultaneous detection associates the ion composition, ion kinetic energies, and electron energies with the same nanoplasma event.
The detected ion composition separates events into populations dominated by \HeIon monocations versus
those dominated by \HeDIon dications, with markedly different ion energies and electron kinetic energies.
The \HeIon-dominated population correlates well with low electron kinetic energies ($< 0.5~\eV$),
whereas the distribution of maximum electron kinetic energies for the \HeDIon-dominated population shows a peak around 1~\eV.
Molecular dynamics simulations averaged over the focal volume suggest that local laser intensity contributes to this separation.
These correlations provide single-event observables for classifying nanoplasmas and
will support real-time shot-to-shot sorting in single-particle coherent diffractive imaging experiments.
\end{abstract}

\section{Introduction}

The interaction of a nanoscale system with intense femtosecond laser fields leads to
the formation of a nanoplasma---a transient, quasi-neutral ensemble of ions and quasi-free electrons.
Nanoplasma formation and the subsequent dynamics have attracted significant attention over the past two decades.
Nanoplasmas efficiently absorb laser energy and have therefore been widely investigated as a source of high-energy particles
\cite{shaoMultikeVElectronGeneration1996, snyderFemtosecondLaserinducedCoulomb1996, trivikramAnomalousIonCharge2013}
and radiation~\cite{mcphersonMultiphotoninducedXrayEmission1994}.
Rare gas clusters, whose mean sizes can be controlled while maintaining low target densities,
provide a versatile and tunable platform for studying nanoplasma formation and relaxation dynamics.
Consequently, such systems have been explored across a broad range of intensities and photon energies,
including the near-infrared (NIR)~\cite{fennelLaserdrivenNonlinearCluster2010, schutteCorrelatedElectronicDecay2015},
extreme ultraviolet (XUV)~\cite{arbeiterIonizationHeatingRaregas2010, mullerIonizationDynamicsXUV2015},
and X-ray domains~\cite{fukuzawaElectronSpectroscopyRaregas2016, hoenerChargeRecombinationSoft2008, kumagaiFollowingBirthNanoplasma2018}.

The mechanisms governing nanoplasma formation depend strongly on the laser wavelength $\lambda$.
In the regime of high ionization potentials and long wavelengths (IR and NIR),
nanoplasma formation is initiated by strong-field tunnel ionization of the constituent atoms
\cite{rose-petruckUltrafastElectronDynamics1997, zheltikovKeldyshParameterPhotoionization2016}.
It is followed, within a few femtoseconds, by avalanche ionization driven primarily by electron-impact ionization (EII)
\cite{saalmannMechanismsClusterIonization2006, schutteLowEnergyElectronEmission2018}.
As ionization progresses, the cluster potential deepens.
Electrons released from their parent atoms or ions can remain confined within the cluster,
a process known as ``inner ionization''~\cite{lastMultielectronIonizationLarge1998}.
Most electrons remain trapped by the collective ionic Coulomb potential,
while some escape directly, leading to ``outer ionization''~\cite{lastElectronNuclearDynamics2004}.
Through collisional energy exchange, some of these electrons gain sufficient energy to escape the cluster,
resulting in thermal electron emission
\cite{shaoMultikeVElectronGeneration1996, kumarappanTwodimensionalEffectsHydrodynamic2002, bostedtFastElectronsMultielectron2010}.
As the nanoplasma expands, the mean kinetic energy of the quasi-free electrons rapidly drops,
and electrons and ions recombine
\cite{fennelHighlyChargedIons2007, schutteRecombinationDynamicsClusters2015, schutteTracingElectronIonRecombination2014}.
In this way, highly excited neutral and ionic Rydberg states are populated,
opening additional relaxation pathways such as fluorescence decay
\cite{sumflethXUVFluorescenceProbe2023, sishodiaXUVFluorescenceProbe2026}
and correlated electronic decay processes similar to interatomic Coulombic decay (ICD)
\cite{schutteCorrelatedElectronicDecay2015, oelzeCorrelatedElectronicDecay2017, ovcharenkoNovelCollectiveAutoionization2014}.

In typical electron or ion spectrometers,
a static electric field is used to project emitted charged particles onto an imaging detector.
As the nanoplasma expands,
this external field can ionize weakly bound electrons in highly excited Rydberg states.
This results in the emission of low-energy electrons
\cite{fennelHighlyChargedIons2007, schutteRareGasClustersIntense2014, medinaSingleshotElectronImaging2021}.
This phenomenon, known as frustrated recombination
\cite{fennelHighlyChargedIons2007},
contributes to the formation of higher ionization
states as final products of the expanding nanoplasma.

Helium (He) nanodroplets serve as an ideal model system for studying laser-driven nanoplasmas.
Because of their simple electronic structure,
the resulting electron spectra are relatively straightforward to interpret.
However, due to the high ionization energy of helium (24.6~\eV), direct ignition of a nanoplasma with
NIR pulses requires a relatively high laser intensity of $\gtrsim 10^{15}~\WcmTwo$.
This limitation can be overcome by doping helium nanodroplets with
a small number of atoms or molecules with lower ionization energies,
such as xenon~\cite{heidenreichEfficiencyDopantinducedIgnition2016}.
The dopants form a cluster at the core of the helium nanodroplets
that facilitates nanoplasma ignition
\cite{mikaberidzeLaserDrivenNanoplasmasDoped2009,krishnanDopantInducedIgnitionHelium2011}.
Alternatively, seed electrons can be introduced via ionization or excitation of the helium itself by XUV pulses
\cite{medinaLonglastingXUVActivation2023, sishodiaXUVFluorescenceProbe2026}.
Previous studies of helium nanoplasmas have primarily relied on the detection of either ions or electrons
\cite{krishnanEvolutionDopantinducedHelium2012, heidenreichEfficiencyDopantinducedIgnition2016,
heidenreichChargingDynamicsDopants2017, kelbgComparisonElectronIon2018, kelbgAugerEmissionCoulomb2019,
kelbgTemporalDevelopmentLaserInduced2020},
mainly yielding ensemble-averaged results over many laser shots and droplet sizes.
Such averaging combines events with different ion compositions and emission energies,
motivating measurements that retain the electron and ion information for each nanoplasma event.

More recently, Medina~\textit{et al.} demonstrated single-shot electron
and ion measurements using a similar experimental setup
\cite{medinaSingleshotElectronImaging2021}.
They found a simple power-law correlation between the energy and
the total counts of emitted electrons.
The present work extends that approach through reconstruction of individual ion
time-of-flight (TOF) traces and classification of the electron spectrum by the detected helium ion species.
This allows us to examine how the electron spectrum correlations depend on whether
\HeIon monocations or \HeDIon dications dominate the accompanying ion signal.

He nanoplasmas have also been used as probes in XUV-pump and NIR-probe
experiments on long-lived ionic and excited states of helium nanodroplets
\cite{medinaLonglastingXUVActivation2023, sishodiaXUVFluorescenceProbe2026}.
In those studies, nanoplasma electron images or nanoplasma fluorescence were recorded as sensitive probes of the time evolution of the XUV-ionized or XUV-excited nanodroplets;
detection of correlated electron and ion spectra would provide more complete information, especially about the initial state of the system (size of the nanodroplet, position within the focal volume) in such experiments.

In this work, we present single-shot, correlated measurements of electrons
and ions emitted from individual helium nanoplasmas driven by intense NIR femtosecond pulses.
Electron velocity-map images (VMI) and ion time-of-flight (TOF) spectra were recorded simultaneously. Analysis of the single-shot ion TOF spectra allows us to classify events into two distinct populations:
those with a reconstructed \HeIon contribution of at least 90\,\%,
and those with a reconstructed \HeDIon contribution of at least 90\,\%.
These populations exhibit markedly different kinetic-energy distributions for both emitted ions and electrons.
The advances over earlier studies of ion and electron spectrometry in intense-laser nanoplasmas are
(i) full electron–ion correlation on a single-shot, single-cluster basis,
(ii) charge-resolved classification of electron emission,
and (iii) comparison with dedicated MD simulations that provide detailed insight into the nanoplasma dynamics.

Section~\ref{sec:experimental_setup} describes the experimental conditions and the synchronization used to record correlated electron VMI images and ion time-of-flight traces.
Section~\ref{sec:results} begins with representative single-shot measurements and the ion TOF analysis.
We then discuss the shape of the average ion charge-state distribution and examine its origin using molecular dynamics simulations.
The subsequent subsections compare ion counts, average ion kinetic energies, electron counts, and maximum electron kinetic energies, and discuss their correlations.
The final subsection presents the correlations between electron and ion counts and kinetic energies.

\section{Experimental setup}
\label{sec:experimental_setup}

\begin{figure}[ht]
\centering
\includegraphics[width=\textwidth]{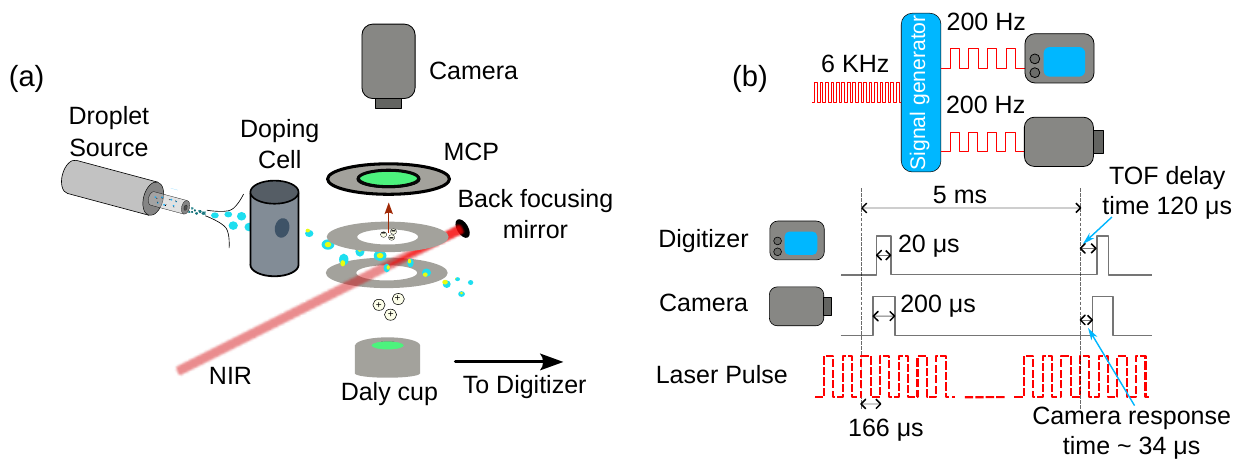}
\caption{
    (a) Schematic of the experimental setup. The He droplets are generated by expanding He gas through a cold continuous nozzle. The droplets are doped with Xe atoms as they pass through a doping cell.
    In the interaction region, a near-infrared (NIR) beam intersects the droplet beam at right angles. The NIR beam is focused using a back-focusing mirror with a focal length of 75~mm. The ions generated by the interaction are collected by a time-of-flight spectrometer containing a Daly cup detector. The signal from the Daly cup is amplified by a photomultiplier and digitized.
    The electrons are detected by a velocity-map imaging (VMI) spectrometer using an MCP detector coupled to a phosphor screen.
    (b) Synchronization scheme for simultaneous single-shot measurement of electron and ion signals.
    The laser trigger is reduced to 200~Hz to accommodate the camera readout rate, and the reduced trigger synchronizes the ion digitizer and the camera recording the electron images.
    The trigger to the digitizer is delayed by 120~$\mu$s, ensuring that both the camera and the digitizer record signals generated by the same laser pulse.
    \label{fig:detector_sync}
}
\end{figure}

Figure~\ref{fig:detector_sync}(a) shows a schematic representation of the experimental setup.
He nanodroplets were produced by continuous supersonic expansion of He gas at
a stagnation pressure of 20 bar through a cryogenic nozzle with a diameter of $5~\mu\mathrm{m}$.
The nozzle was cooled to a temperature between 10~K and 12~K.
For these nozzle temperatures, the estimated average nanodroplet size before doping
ranges from $\mean{N}=10^4$ to $10^5$ He atoms per droplet.
These estimates were obtained using the HeNDS program~\cite{rastonHeNDSProgramCalculating2021}
and by comparison with literature values~\cite{toenniesSuperfluidHeliumDroplets2004}.
The nanodroplets were doped in a separate vacuum chamber containing a gas doping cell with a length of 18~mm.
Xenon gas was admitted into the doping cell through a fine dosing valve,
resulting in a pressure of $3.6\times10^{-5}$~mbar in the doping chamber.
We estimate the mean number of xenon atoms incorporated into the droplets to range from $\mean{N_d}=21$ to $75$
by simulating the doping process as described in Ref.~\cite{deFragmentationWaterClusters2024}
using the open-source software HeDopant~\cite{sishodiaHeDopantProgramSimulation2026}.
Evaporation of He atoms during the doping process reduces the nanodroplet size.
Taking this reduction into account, we estimate that the doped nanodroplet distributions have mean sizes of
$\mathrm{He}_{26468}\mathrm{Xe}_{75}$,
$\mathrm{He}_{6812}\mathrm{Xe}_{32}$, and
$\mathrm{He}_{3792}\mathrm{Xe}_{22}$ for nozzle temperatures of 10~K, 11~K, and 12~K, respectively.
Intense near-infrared (NIR) pulses were generated using a commercial femtosecond laser system operating at a repetition rate of 6~kHz.
The laser beam was focused inside the interaction chamber using a back-focusing mirror with a focal length of 75~mm.
Pulse energies of up to $270~\mu$J resulted in peak intensities up to $10^{15}~\WcmTwo$,
as inferred from the charge-state distribution measured by strong-field ionization of
Xe atoms~\cite{augstLaserIonizationNoble1991} (see Fig.~S1).

Ions were detected using an ion TOF spectrometer equipped with
a Daly-type scintillation detector~\cite{dalyScintillationTypeMass1960},
whereas electrons were imaged onto a microchannel plate (MCP) detector coupled to a phosphor screen.
To extract the electrons and ions, an electric field of $127~\mathrm{V/cm}$
was applied across the interaction region.
The ion signals from the Daly detector were digitized using
a fast ADC data card,
and a CCD camera captured the electron images on the phosphor screen.
A delay generator triggered by the laser pulse provided a common 200~Hz trigger
for both detectors, as illustrated in Fig.~\ref{fig:detector_sync}.
The synchronization trigger rate was limited by the response time and acquisition time
of the CCD camera.
The digitizer trigger was delayed by 120~$\mu$s so that each recorded ion TOF trace
and electron VMI corresponded to the same laser pulse.
This synchronization allowed us to correlate the electron and ion signals
for individual nanoplasma events.
For a given mean droplet size, 50,000 single shots were typically recorded,
of which only about 600 contained a nanoplasma event,
identified by a bright circular feature in the electron VMI
with a radius of $R_{\mathrm{max}}\geq5$~pixels (see the Supplementary Material).
Assuming independent nanoplasma hits with Poisson statistics,
a hit probability $p\approx0.01$ corresponds to
a mean $\lambda=-\ln(1-p)\approx0.010$ per shot.
The probability of two or more hits is then
$1-e^{-\lambda}(1+\lambda)\approx5\times10^{-5}$ per shot,
or approximately 0.5\,\% of hit-containing shots.
Thus, the measurements were performed predominantly in a single-shot single-cluster regime,
with $\approx 0.5\%$ of hit-containing shots expected to contain more than one nanoplasma event.
Separately, from the pressure contrast between the He droplet beam and the background in the beam dump
($\Delta P_{\mathrm{dump}} \approx 4\times10^{-9}\,\mathrm{mbar}$),
we can estimate the He particle density to be
$\approx 2.6\times10^{8}~\mathrm{cm}^{-3}$ in the interaction volume.
For an interaction volume of $\approx5\times10^3~\mu\mathrm{m}^3$, this corresponds
to approximately 1.3 He atoms on average.

\section{Results and discussion}
\label{sec:results}

\subsection{Single-shot ion TOF analysis}
\label{sec:single_shot_analysis}

\begin{figure}[ht]
\centering
\includegraphics[width=\textwidth]{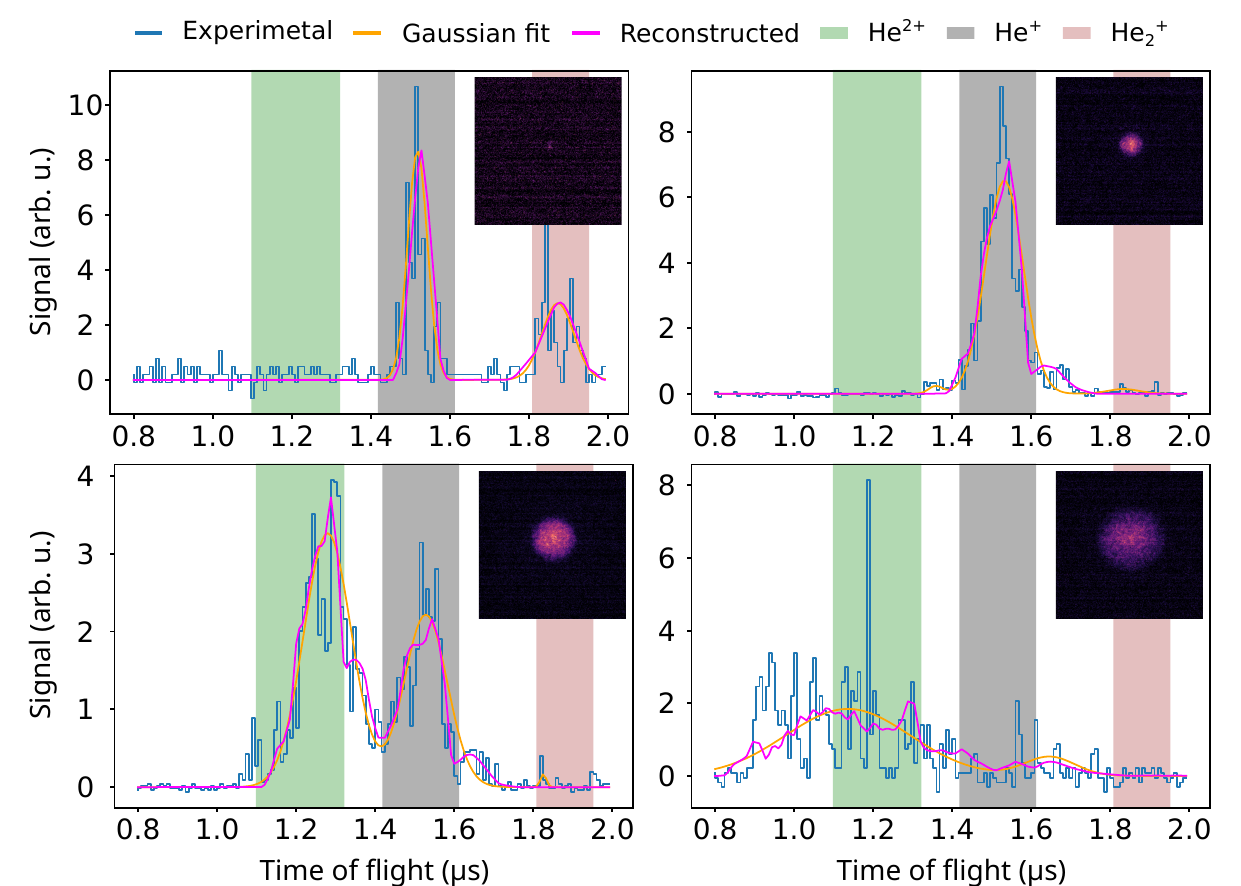}
\caption{
    Sample single-shot ion TOF traces (blue) and electron images (insets).
    The Gaussian fits and spectra reconstructed using SIMION simulations are shown in orange and magenta, respectively.
    The expected TOF intervals for the individual ion species are highlighted by colored areas.
}
\label{fig:sample_shots}
\end{figure}

Figure~\ref{fig:sample_shots} shows representative single-shot measurements.
Each panel shows the TOF spectrum of a single nanoplasma event in blue; the corresponding electron images are shown in the insets.
For each nanoplasma event, the expected TOF regions for \HeDIon, \HeIon, and \HeTwoIon ions are highlighted in green, gray, and red, respectively.
To obtain information about the kinetic-energy distributions of these ions,
we fitted the TOF spectra with multi-peak Gaussian functions, shown in orange.
We then used a technique based on earlier work by Ngai~\textit{et al.}~\cite{ngaiMethodKineticEnergy2024}
to reconstruct the ion TOF traces from simulations and determine the kinetic energies of \HeIon, \HeDIon, and \HeTwoIon ions.

We used SIMION~\cite{dahlSimionPersonalComputer2000} to simulate ion trajectories in the spectrometer
and generate a set of basis functions for different initial ion kinetic energies and mass-to-charge ratios.
To prevent the basis functions from fitting the substantial shot noise in many single-shot TOF spectra,
we first represented each measured TOF trace by a smooth multi-peak Gaussian fit.
We then fitted a linear combination of the SIMION basis functions to that smooth trace.
The fitted basis-function weights determine the reconstructed kinetic energy distribution,
while the Gaussian fit reduces sensitivity to fluctuations in the raw trace.
The TOF spectra reconstructed using the basis functions are shown in magenta in Fig.~\ref{fig:sample_shots}.
From this reconstruction, we infer the relative contributions of the individual ion species,
$W_{\mathrm{ion}}$, and calculate the average ion kinetic energy, \Ei.
Together, these quantities characterize the ion composition and emission energy
of each nanoplasma event.
Further details of the analysis are provided in the Supplementary Material.

\subsection{Average ion charge-state distribution}
\label{sec:charge_distribution}

\begin{figure}[ht]
\centering
\includegraphics[width=\textwidth]{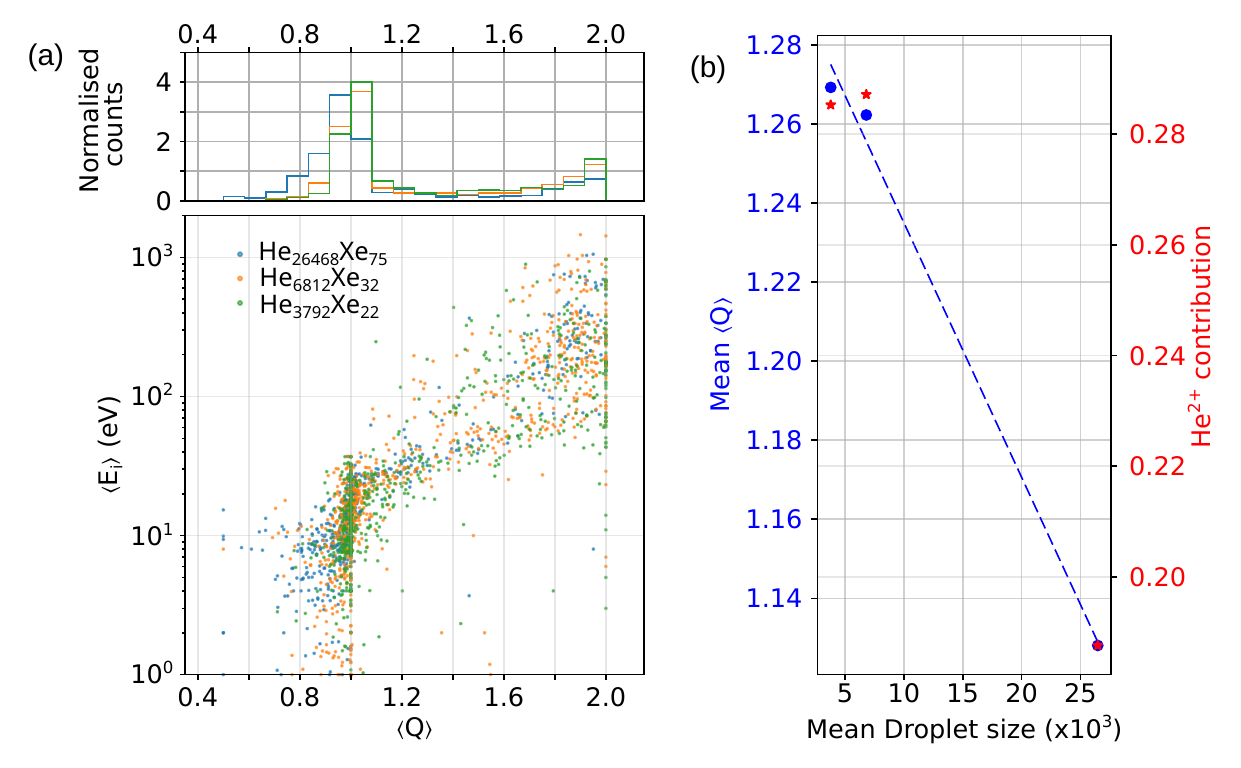}
\caption{
    Panel (a) shows the average charge of emitted ions $\mean{Q}$ versus
    the average kinetic energy $\Ei$ for individual nanoplasma events, represented by colored dots.
    Green, orange, and blue represent three measurement sets for different He nanodroplet sizes.
    $\mean{Q}$ is calculated using Eq.~\ref{eq:avg_charge}.
    The top panel shows a normalized histogram of $\mean{Q}$ values obtained by summing the data over $\Ei$.
    Panel (b) shows $\mean{Q}$ and the mean contribution of \HeDIon ions for different mean droplet sizes.
    The dashed red and blue lines are guides to the eye.
    \label{fig:ion_charge}
}
\end{figure}

For each nanoplasma event, we calculate the average ion charge from the fit results,
\begin{equation}
    \mean{Q} = 2.0 \, W_{\HeDIon} + 1.0 \, W_{\HeIon} + 0.5 \, W_{\HeTwoIon},
    \label{eq:avg_charge}
\end{equation}
where $W_{\mathrm{ion}}$ is the normalized contribution of the specified helium ion
species to the reconstructed TOF trace,
with the three contributions summing to one.
The coefficients 2.0 for \HeDIon, 1.0 for \HeIon, and
0.5 for \HeTwoIon represent the charge per He atom.
Thus, events containing only \HeIon or only \HeDIon contributions have
$\mean{Q}=1.0$ or $\mean{Q}=2.0$, respectively, while a pure \HeTwoIon contribution gives
$\mean{Q}=0.5$.
The average ion charge therefore provides a compact description of the detected ion composition.
Similarly, the average kinetic energy of the emitted ions,
$\Ei$, is calculated for each nanoplasma event as
the weighted average of the ion kinetic energies obtained from the reconstructed spectra.

The resulting average ion kinetic energies $\Ei$ versus $\mean{Q}$ are shown in Figure~\ref{fig:ion_charge}(a)
for three measurement sets with different mean He nanodroplet sizes.
Each dot represents a single nanoplasma event identified among the 50,000 laser shots
recorded for that measurement set.
The top panel is a histogram of $\mean{Q}$ values obtained by summing the data over $\Ei$.
The correlation map and histogram of $\mean{Q}$ show the following trends:
\begin{itemize}
    \item The distribution of the average charge of the emitted ions,
    $\mean{Q}$, is U-shaped, with peaks at $\mean{Q}=1.0$ (TOF spectra containing mostly \HeIon ions)
    and $\mean{Q}=2.0$ (TOF spectra containing mostly \HeDIon ions).
    \item Nanoplasma events with $\mean{Q}=2.0$ are less abundant than those with $\mean{Q}=1.0$.
    \item A higher average charge $\mean{Q}$ correlates with a higher average kinetic energy of ions, $\Ei$.
    \item Larger He nanodroplets (blue dots) have a lower average charge than smaller nanodroplets (green dots).
    The trend in the mean values is shown in Fig.~\ref{fig:ion_charge}(b).
\end{itemize}

\subsection{Molecular dynamics simulations}
\label{sec:focal_volume_md}

The U-shaped distribution of average ion charge, $\mean{Q}$,
can be explained by two nonlinear dependencies that affect the detected signals:

(i) The laser focus is described by a Gaussian beam profile. The corresponding focal-volume intensity distribution is $g(I)$,
which is highly nonlinear and gives much greater weight to lower intensities
\cite{heidenreichKineticEnergyDistribution2011}:
\begin{equation}
    g(I) = \frac{\pi w_0^2 z_r}{V(I_{min})} \frac{\zeta(I)^{1/2} + \frac{1}{3}\zeta(I)^{3/2}}{I},
    \label{eq:spatial_density}
\end{equation}
where
\begin{equation}
    V(I) = \frac{4}{3}\pi w_0^2z_r \left( \zeta(I)^{1/2} + \frac{1}{6}\zeta(I)^{3/2} - \text{tan}^{-1}\zeta(I)^{1/2} \right)
\end{equation}
and
\begin{equation}
    \zeta(I) = \frac{I_{max} - I}{I}.
\end{equation}
Here, $w_0$ is the radius of the Gaussian beam at the focus, $z_r=\pi w_0^2/\lambda$ is the Rayleigh range,
$I_{max}$ is the peak intensity, and $I_{min}$ is the minimum pulse peak intensity required to produce a nanoplasma.

(ii) The formation of high charge states in a nanoplasma is itself highly nonlinear with respect to the laser intensity.
The interplay between these two nonlinearities gives rise to the U-shaped charge-state distribution.

\begin{figure}
    \centering
    \includegraphics[width=0.8\textwidth]{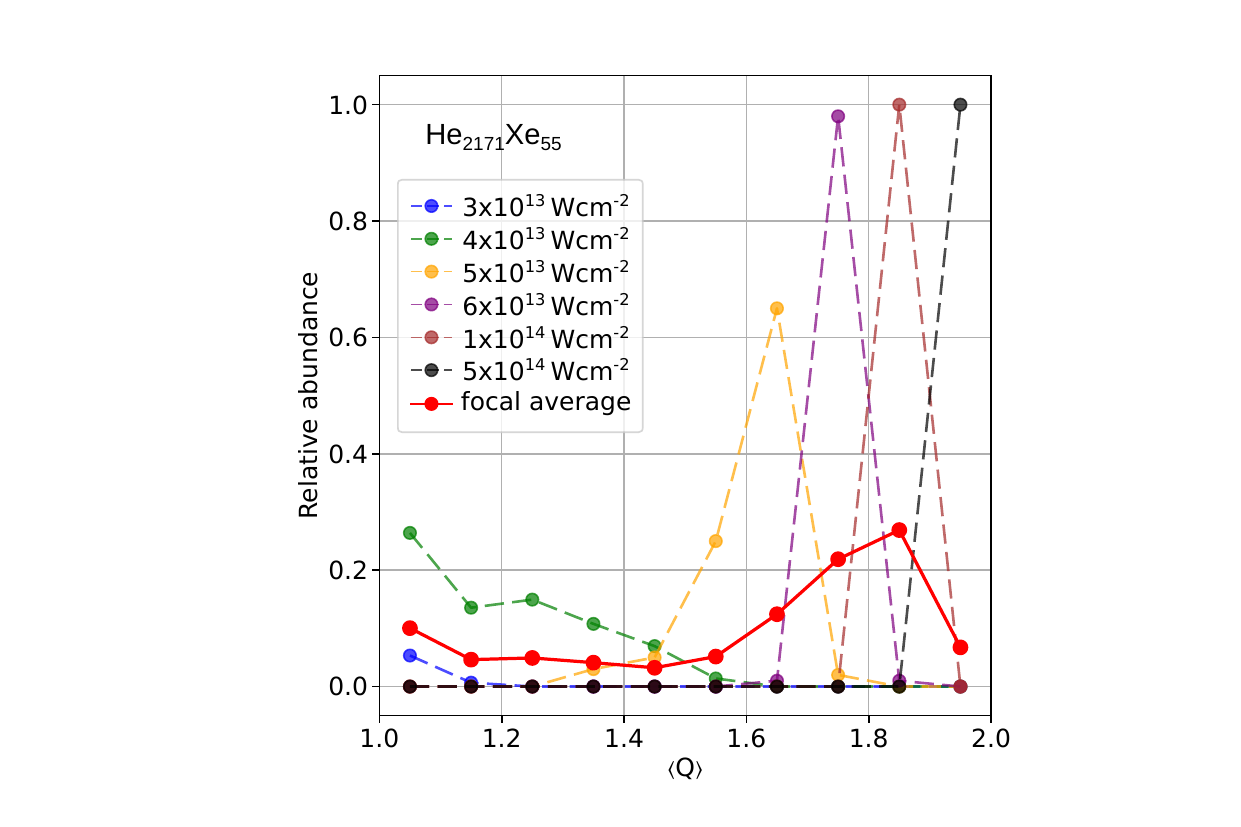}
    \caption{
        The average charge of a He droplet (He$_{2171}$Xe$_{55}$) at different laser intensities, calculated using MD simulations.
        The results for different laser intensities were averaged over the focal volume according to
        the intensity distribution given by Eq.~\ref{eq:spatial_density}.
        The distribution averaged over the focal volume is shown in red.
        \label{fig:md}
    }
\end{figure}

To assess this interpretation, we carried out molecular dynamics (MD) simulations for a model cluster of He$_{2171}$Xe$_{55}$.
In these simulations, all atoms and nanoplasma electrons were propagated classically.
Calculations were carried out for selected peak intensities within the focal spot,
using a photon energy of 1.55~\eV and a Gaussian temporal intensity envelope
with a full width at half maximum (FWHM) of 130~fs.
For each intensity $I_M \geq 5\times10^{13} \WcmTwo$,
we generated an ensemble of 100 trajectories with different initial conditions.
For sample intensities $I_M \leq 5\times10^{13} \WcmTwo$, we generated an ensemble of 300 trajectories.
To account for three-body ion–electron recombination (TBR),
trajectories were propagated for 2~ps, except at $I_M=4\times10^{13}~\WcmTwo$,
where 3~ps was used to account for the slower nanoplasma expansion.
The mean charge state $\mean{Q}$ of all He ions in the droplet was evaluated
at the end of each trajectory.

Figure~\ref{fig:md} presents the resulting distributions of $\mean{Q}$ for the different laser intensities (dashed lines).
The results for different laser intensities were averaged over the 3D Gaussian profile of the NIR pulse,
as defined in Eq.~\ref{eq:spatial_density}.
A more detailed description of the MD methodology is provided in the Supplementary Material (SM).
The focal-volume average gives a U-shaped distribution, with higher charges
($\mean{Q}\approx2$) more abundant, whereas the experimental distribution is
weighted toward $\mean{Q}\approx1$ (Fig.~\ref{fig:ion_charge}).
The simulations therefore reproduce the shape of the distribution qualitatively,
but give different relative abundances of the two populations.

Several differences between the simulations and the experiment can affect this comparison.
The simulations evaluate the charge state of all He ions at the end of each trajectory, whereas the
experimental average ion charge is constructed from detected TOF contributions.
Figure~\ref{fig:ion_charge}(a) shows that events with $\mean{Q}=2.0$
have much higher average ion kinetic energies than events with $\mean{Q}=1.0$.
The transmission of the TOF spectrometer depends on the ion kinetic energy and decreases
to approximately 10\% for ion kinetic energies exceeding 250~\eV (see Fig.~S2).
Furthermore, the experiment uses much larger droplets than those in the MD simulations; large helium nanodroplets are prone to partial ionization and to enhanced electron-ion recombination. This is suggested by Fig.~\ref{fig:ion_charge}(b),
which shows a trend toward lower $\mean{Q}$ values for larger droplets.
Partial ionization results in the emission of \HeIon and He$_2^+$ ions, leading to $\mean{Q} \lesssim1.0$.
A lower effective peak intensity than the estimated value, e.g., due to deviations from an ideal Gaussian beam profile, also increases the relative contribution
of the low-charge population.
Thus, the energy-selective transmission of the ion TOF spectrometer, combined with partial ignition of large nanodroplets and a lower-than-estimated laser intensity, can explain the lower abundance of nanoplasma events with $\mean{Q} = 2.0$ compared to the MD simulations.

\subsection{Ion counts and kinetic energies}
\label{sec:ion_correlations}

\begin{figure}[ht]
\centering
\includegraphics[width=\textwidth]{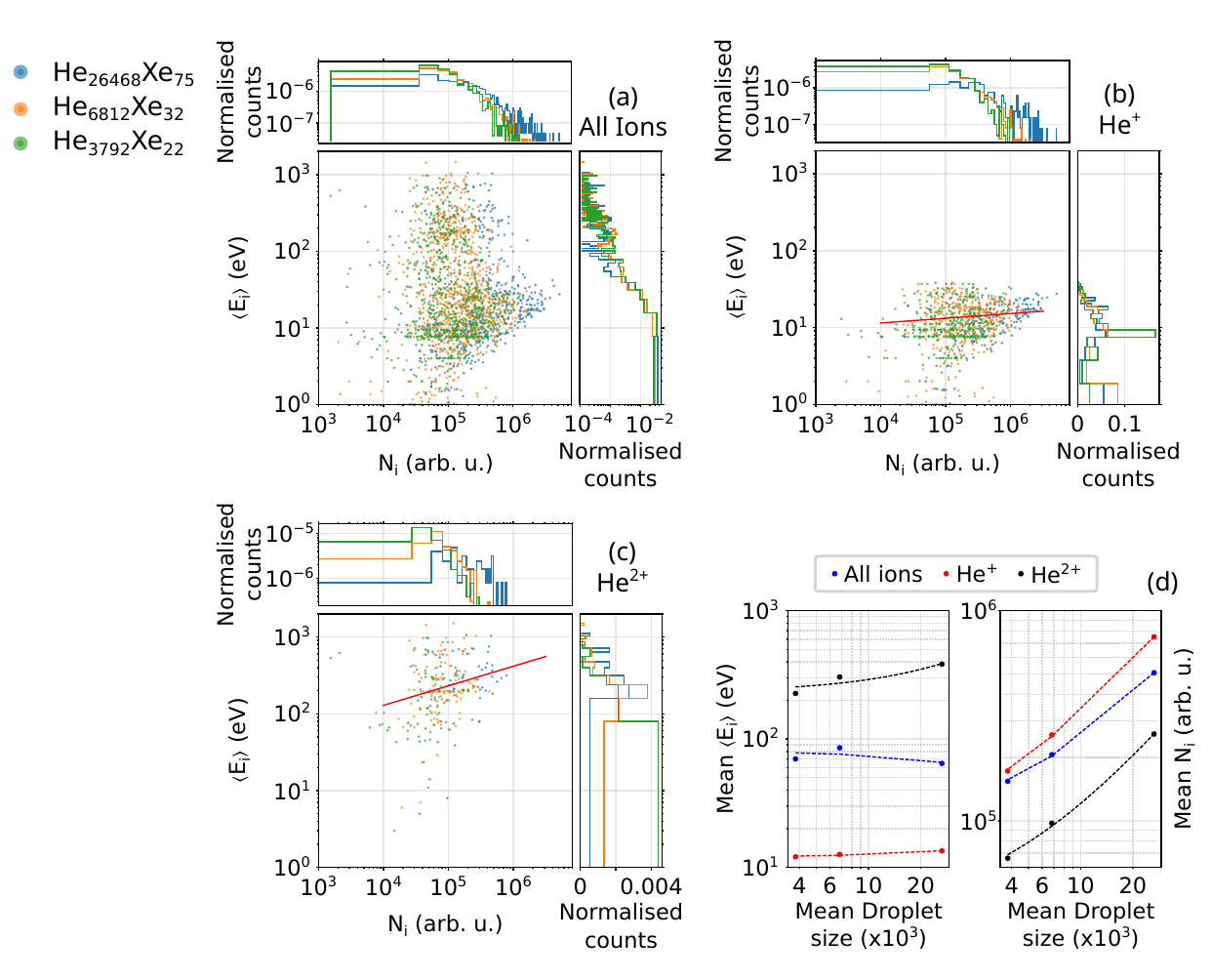}
\caption{
   Panel (a) plots the average ion kinetic energy ($\Ei$) versus total ion counts ($\Ni$) for individual nanoplasma events (colored dots).
   Green, orange, and blue indicate three He nanodroplet sizes (listed in the legend).
   Panels (b) and (c) show $\Ei$–$\Ni$ correlations for events dominated by \HeIon and \HeDIon, respectively.
   Panels (a)–(c) each include two histograms: the top histogram for $\Ni$ and the right histogram for $\Ei$ at different droplet sizes.
    The correlation is fitted with a power law $\Ei \propto \Ni^n$ in (b) and (c).
    Panel (d) shows the mean ion kinetic energies (left) and mean integrated ion signals (right) versus the mean droplet size for all events and the two selected populations. Dashed lines are guides to the eye.
    \label{fig:ion_dist}
}
\end{figure}

The U-shaped average charge-state distribution motivates separate analyses of
(i) events dominated by \HeDIon ions ($\mean{Q}\approx2$) and
(ii) events dominated by \HeIon ions ($\mean{Q}\approx1$).
This classification allows us to compare the ion yields and kinetic energies
of the two populations.
Figure~\ref{fig:ion_dist}(a)–(c) shows the correlation between the total ion counts $\Ni$ and
the average ion kinetic energy $\Ei$ for individual nanoplasma events at three mean He droplet sizes.
Each dot represents a single nanoplasma event, with $\Ni$ calculated by integrating
its ion TOF trace.
The top and right histograms show the normalized distributions of $\Ni$ and $\Ei$, respectively.
Panel (a) includes all nanoplasma events.
The distribution of $\Ni$ spans approximately three orders of magnitude,
with the measurement set having the largest mean droplets (blue dots) extending to the highest ion signals.
The corresponding $\Ei$ values also span approximately three orders of magnitude,
with most events concentrated around $\Ei \approx 20~\eV$.
A subset of events has substantially higher average ion kinetic energies,
reaching several hundred \eV.
Figure~\ref{fig:ion_dist}(b) shows events dominated by \HeIon monocations,
with $W_{\HeIon}\geq0.9$;
Fig.~\ref{fig:ion_dist}(c) shows events dominated by \HeDIon dications, with $W_{\HeDIon}\geq0.9$.

The correlations for the two selected populations are fitted empirically with
$\Ei=a\,\Ni^n$.
The fitted exponents are $n=0.06$ for \HeIon-dominated events and $n=0.2$ for \HeDIon-dominated events,
indicating a much weaker fitted dependence for the former population.
Figure~\ref{fig:ion_dist}(d) shows the mean \Ei and mean \Ni as functions of mean droplet size.
The \HeIon-dominated events have larger mean detected \Ni than
the \HeDIon-dominated events at all three mean droplet sizes.
Energy-dependent transmission of the TOF spectrometer may contribute to this difference.
The \HeDIon-dominated events predominantly have high average ion kinetic energies ($\gtrsim100~\eV$).
Simulations for \HeDIon ions show that the transmission decreases strongly with kinetic energy,
reaching approximately 10\,\% above 250~\eV (Fig.~S2).
Consequently, the measured ion yield for nanoplasma events dominated by \HeDIon ions is significantly diminished.
The observed difference in ion yields therefore reflects both the emission from the
nanoplasma and the energy-dependent transmission of the spectrometer.

The \HeDIon- and \HeIon-dominated populations occupy different regions of the \Ei--\Ni correlation map.
The MD simulations suggest that higher local laser intensities favor higher final ion charge states.
The higher ion kinetic energies associated with \HeDIon-dominated events are consistent with stronger ion acceleration during Coulomb explosion.
Figure~\ref{fig:ion_dist}(d) shows that the mean \Ei is higher for \HeDIon-dominated events than for \HeIon-dominated events at all three mean droplet sizes.
Although the mean values of \Ei for both selected populations increase with mean droplet size,
the all-event mean is lower for the largest droplets than for the intermediate size;
it does not decrease monotonically across all three sizes.
This is consistent with the trend toward lower average detected charge in Fig.~\ref{fig:ion_charge}(b).
This arises because nanoplasma events with $\mean{Q} \lesssim 1.0$ are more abundant for the largest
droplet sizes.

The sensitivity of the detected ion composition to local laser intensity suggests its potential use as a statistical diagnostic of the overlap between the cluster beam and the laser.
However, the dominant species or $\mean{Q}$ does not uniquely determine a droplet's position within the focus:
droplet size, ionization dynamics, and spectrometer transmission also affect the observables.
Combining the detected ion composition with $\Ni$ may help constrain the contributions of droplet size and local intensity
and can help distinguish a large cluster outside the most intense part of the laser from a smaller cluster inside it.
Determining $\Ei$ requires the two-stage TOF reconstruction involving human intervention,
whereas integrated ion signals and approximate species contributions could potentially be extracted with less processing.
These observables are therefore well suited for integration with online or near-online data analysis pipelines used in the community
\cite{thayerDataSystemsLinac2017, maroneOntheflyDataPostprocessing2017}.

\subsection{Electron counts and kinetic energies}
\label{sec:electron_correlations}

\begin{figure}[ht]
\centering
\includegraphics[width=\textwidth]{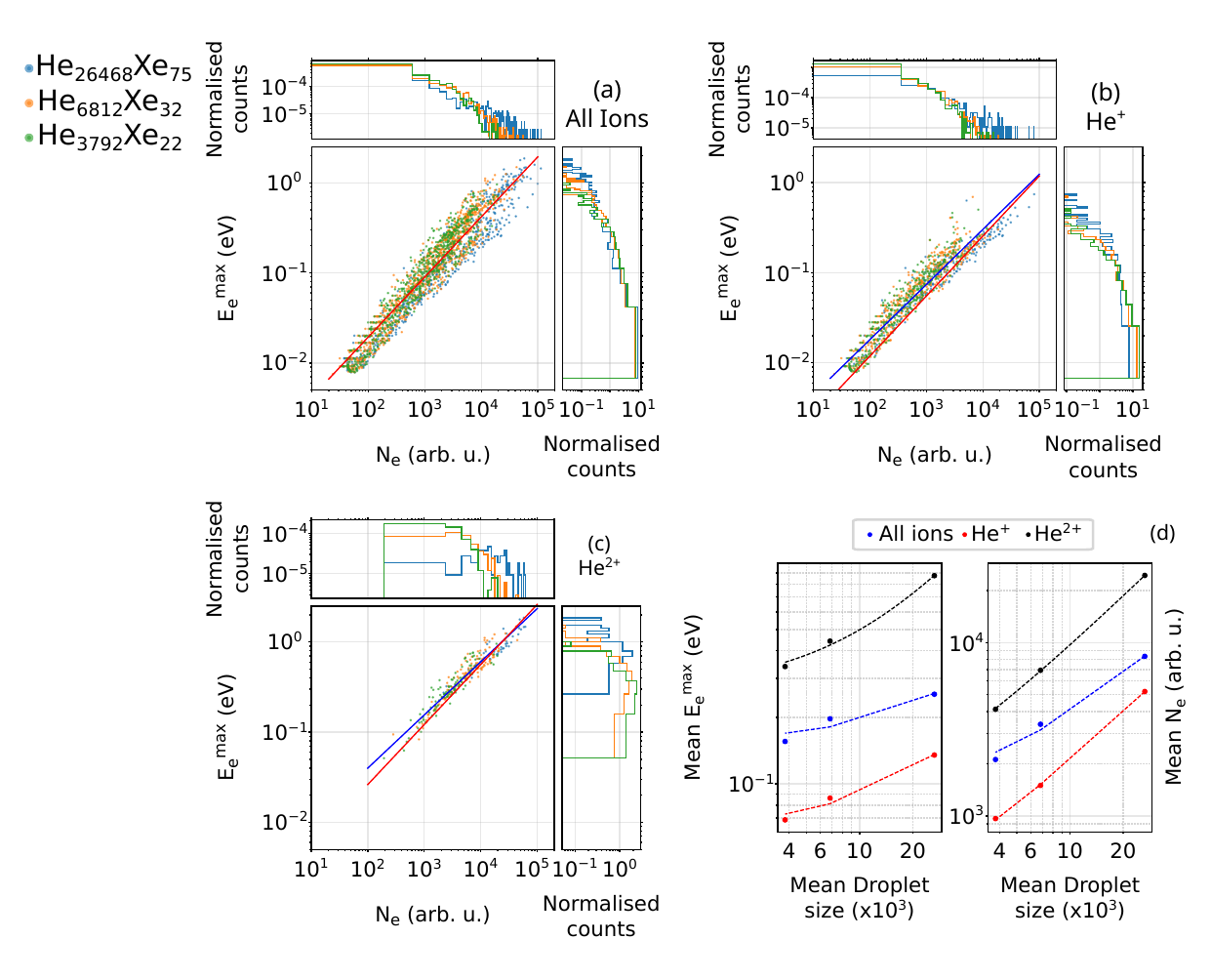}
\caption{
    Panel (a) shows the maximum electron kinetic energy ($\Ee$)
    versus total electron counts ($\Ne$) for individual nanoplasma events (colored dots).
    Green, orange, and blue denote three measurement sets for different He nanodroplet sizes, as listed in the legend.
    Panels (b) and (c) show $\Ee$ versus $\Ne$ for \HeIon-dominated ($W_{\HeIon}\geq0.9$)
    and \HeDIon-dominated ($W_{\HeDIon}\geq0.9$) events, respectively.
    Panels (a)–(c) also include histograms: $\Ne$ on top and $\Ee$ on the right.
    The overall correlation is fitted by $\Ee \propto \Ne^{2/3}$ (red line).
    Fits with a free exponent, $\Ee\propto\Ne^n$, for the two selected populations are shown in blue.
    Panel (d) presents the mean $\Ee$ and $\Ne$ versus mean droplet size.
    \label{fig:eke}
}
\end{figure}

The electron VMIs, shown in the insets of Fig.~\ref{fig:sample_shots},
were analyzed to obtain the maximum kinetic energy of the emitted electrons, $\Ee$,
and the total integrated electron counts, $\Ne$.
First, a Fourier filter was used to remove high-frequency noise from the raw electron VMI images.
The radius $R_{\mathrm{max}}$ of the bright circular feature was converted to an electron velocity
and then to the maximum kinetic energy, $\Ee$, for each nanoplasma event.
Each event thus contributes one $\Ee$ value to the distributions discussed below.
The total electron counts ($\Ne$) were calculated by integrating the image intensity
within $r \leq R_{\mathrm{max}}$.
Figure~\ref{fig:eke}(a)–(c) shows the correlation between $\Ee$ and $\Ne$ for individual
nanoplasma events at three mean He droplet sizes.
Each dot represents one nanoplasma event, and the top and right histograms show
the normalized distributions of $\Ne$ and $\Ee$, respectively.
Panel (a) includes all nanoplasma events.
The correlation is fitted with a simple power law $\Ee\propto \Ne^{2/3}$,
using the formalism developed by Medina~\textit{et al.}~\cite{medinaSingleshotElectronImaging2021}.
As a simple size-scaling argument, at fixed charge per atom and atomic density,
a Coulomb energy scale varies as $N/R$, with $R\propto N^{1/3}$, giving $N^{2/3}$
\cite{islamKineticEnergyIons2006}.
Applying this scaling to $\Ne$ assumes that the detected signal from the central feature
tracks cluster size.
Using the correlated ion measurements and the U-shaped average charge distribution (Fig.~\ref{fig:ion_charge}),
we plotted charge-selected nanoplasma events in Fig.~\ref{fig:eke}(b) and (c).
For each population, we compare a fit with the fixed exponent $2/3$ (red)
with a fit using a free exponent, $\Ee\propto\Ne^n$ (blue).
This comparison shows how the charge-selected electron correlations deviate
from the scaling used for the full ensemble.
Figure~\ref{fig:eke}(d) shows the average $\Ee$ and $\Ne$ as a function of droplet size.
The distributions of $\Ee$ differ markedly between the \HeIon- and \HeDIon-dominated events.
The \HeIon-dominated population is concentrated at lower electron energies,
whereas the \HeDIon-dominated population extends to higher energies.
The mean $\Ne$ depends linearly on the mean droplet size.
For \HeDIon-dominated nanoplasma events, the mean $\Ne$ is approximately four times
as large as for \HeIon-dominated events.

\begin{figure}[ht]
\centering
\includegraphics[width=0.7\textwidth]{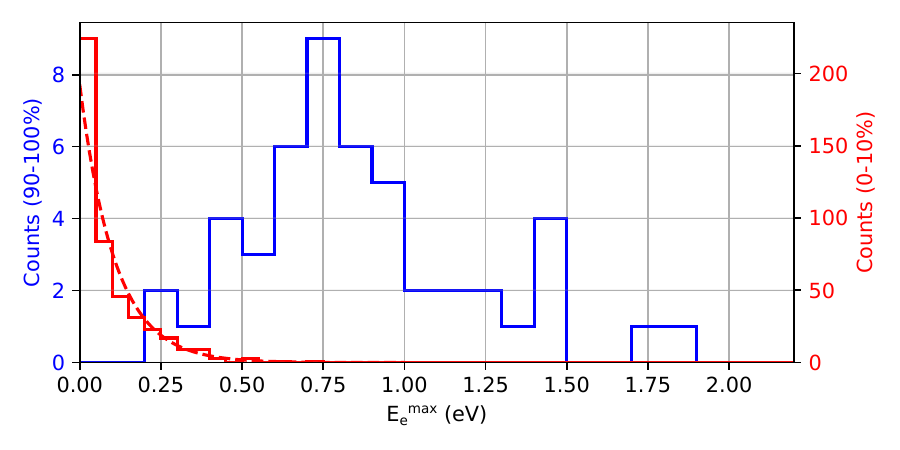}
\caption{
    Histogram of the maximum electron kinetic energy, \Ee, for He droplets of size He$_{26468}$Xe$_{75}$.
    Nanoplasma events are selected based on the contribution from doubly charged \HeDIon ions.
    The blue line shows events with a contribution $\geq 90~\%$,
    and the red line shows events with a contribution $\leq 10~\%$.
    The dashed red line is an exponential fit to the \Ee distribution for contributions $\leq 10~\%$.
    \label{fig:he2+_filter}
}
\end{figure}

Figure~\ref{fig:he2+_filter} highlights the different \Ee distributions for high and low
\HeDIon contributions in the He$_{26468}$Xe$_{75}$ measurement set.
The blue histogram selects events with $W_{\HeDIon}\geq0.9$,
while the red histogram selects events with $W_{\HeDIon}\leq0.1$.
The low-\HeDIon population ($W_{\HeDIon} \leq 0.1$) exhibits an exponential electron energy distribution,
with most events below 100~meV.
Ignition by the leading edge of the femtosecond laser pulse and subsequent avalanche
ionization produce a deep Coulombic potential that traps most of the released electrons.
The trapped electrons and parent ions form the nanoplasma.
As the nanoplasma expands and cools, these electrons recombine with the parent ions
to form highly excited Rydberg atoms
\cite{schutteRecombinationDynamicsClusters2015, heidenreichChargingDynamicsDopants2017, komarHighlyChargedRydberg2018}.
One possible source of electrons with low kinetic energies is
the ionization of these Rydberg atoms by the electric field used in the electron VMI spectrometer
\cite{fennelHighlyChargedIons2007, schutteRareGasClustersIntense2014,
schutteRecombinationDynamicsClusters2015, komarHighlyChargedRydberg2018}.
The electron kinetic energy for this type of process depends on the electric fields
of the VMI~\cite{schutteRareGasClustersIntense2014, medinaSingleshotElectronImaging2021}.
Another possible source is evaporation of the nanoplasma electrons
during the expansion and cooling phase of the nanoplasma dynamics.
This thermal emission of quasi-free nanoplasma electrons
gives rise to an exponential distribution, with the majority of electrons
having kinetic energies of a few meV
\cite{arbeiterIonizationHeatingRaregas2010,
kelbgComparisonElectronIon2018}.

Events selected by $W_{\HeDIon} \geq 0.9$ show a peak around 0.75~\eV in the $\Ee$ histogram.
Nanoplasma events that lead to fully ionized atoms occur
at high laser intensities (Fig.~\ref{fig:md}).
The nanoplasma formed during these events would be denser;
thus, the electron cloud would experience higher Coulomb repulsion
as it expands after leaving the cluster potential.
This might result in higher kinetic energy in the case of
$W_{\HeDIon} \geq 0.9$ compared to $W_{\HeDIon} \leq 0.1$.
Correlated electronic decay between excited nanoplasma ions offers another possible
pathway for producing a specific kinetic energy peak
\cite{schutteObservationCorrelatedElectronic2015}.
This can generate electrons with a specific kinetic energy, such
as $\sim 0.75~\eV$, in coincidence with \HeDIon ions.
Using a field-free electron detection method, Kelbg~\textit{et~al.} observed a similar peak
in the averaged electron kinetic-energy spectrum of He nanoplasmas
\cite{kelbgComparisonElectronIon2018}.
As the detection method was field-free, re-ionization of Rydberg states did not occur.
They measured the average electron kinetic energy spectrum with
a much larger dynamic range than that presented here,
which allowed them to detect even weak peaks.
They detected two different electron kinetic energy distributions,
including an exponential distribution with electron kinetic energies $< 0.5~\eV$,
which they attributed to the thermal evaporation of electrons from
the nanoplasma.
They also detected peaks at specific energies after a time delay and
attributed these peaks to correlated decay.
Our charge-selected analysis separates nanoplasma events with different \Ee distributions
and associates these distributions with the accompanying ion composition.
Correlated decay would also be expected to produce an electron with a higher kinetic energy.
Such electrons were not detected, owing to the VMI spectrometer's limited energy
acceptance and the much lower yield of high-energy electrons.
The present measurements therefore leave the contribution of this pathway unresolved.
Sch{\"u}tte~\textit{et~al.} have detected low-energy
electron emission ($< 4~\eV$) using a field-free spectrometer,
which allowed them to avoid the ionization of Rydberg states by the detector field
\cite{schutteLowEnergyElectronEmission2018}.
They attributed these low-energy electrons to the autoionization of the excited cluster nanoplasma.

\subsection{Electron--ion correlations}
\label{sec:electron_ion_correlations}

\begin{figure}[ht]
\centering
\includegraphics[width=\textwidth]{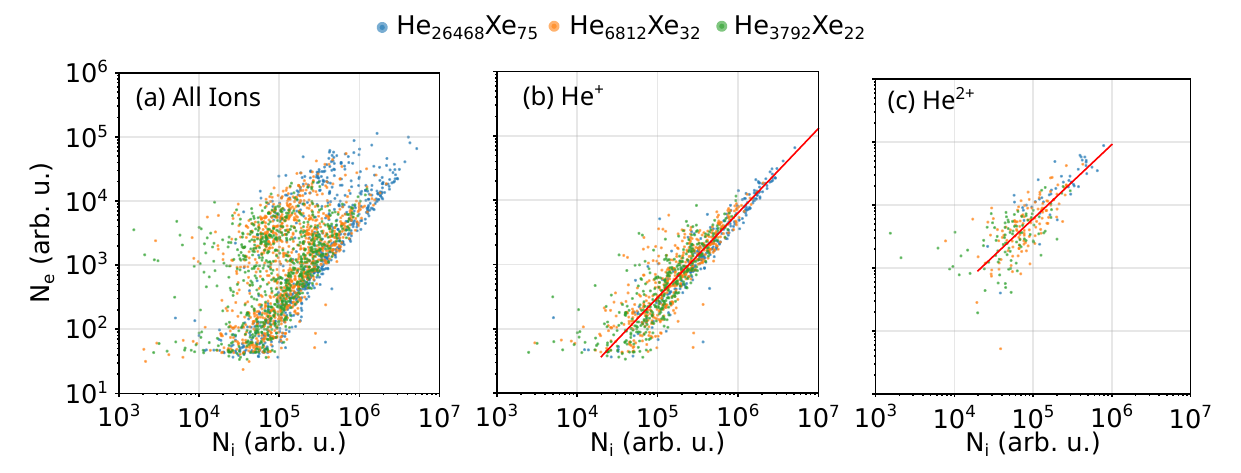}
\caption{
    Panel (a) shows $\Ne$ versus $\Ni$ for individual nanoplasma events (colored dots).
    Blue, orange, and green denote the He$_{26468}$Xe$_{75}$, He$_{6812}$Xe$_{32}$,
    and He$_{3792}$Xe$_{22}$ measurement sets, respectively.
    Panels (b) and (c) select \HeIon-dominated ($W_{\HeIon}\geq0.9$)
    and \HeDIon-dominated ($W_{\HeDIon}\geq0.9$) events, respectively.
    Red lines show power-law fits, $\Ne\propto\Ni^n$, to the selected populations.
    \label{fig:ion_ele_counts}
}
\end{figure}

Figure~\ref{fig:ion_ele_counts} shows the correlation between the detected electron and ion counts
for individual nanoplasma events for three mean He droplet sizes.
Each dot represents a single nanoplasma event.
Panel (a) includes all events, while panels (b) and (c) select events dominated by
\HeIon and \HeDIon, respectively.
The correlations for the selected populations are fitted with a simple power law,
$\Ne \propto \Ni^n$.
At fixed ion composition, complete detection of the released electrons and ions
would give a linear relationship from charge balance.
For \HeDIon, the best fit is $\Ne \propto \Ni^{1.1}$, while for \HeIon it is $\Ne \propto \Ni^{1.3}$.

\begin{figure}[ht]
\centering
\includegraphics[width=\textwidth]{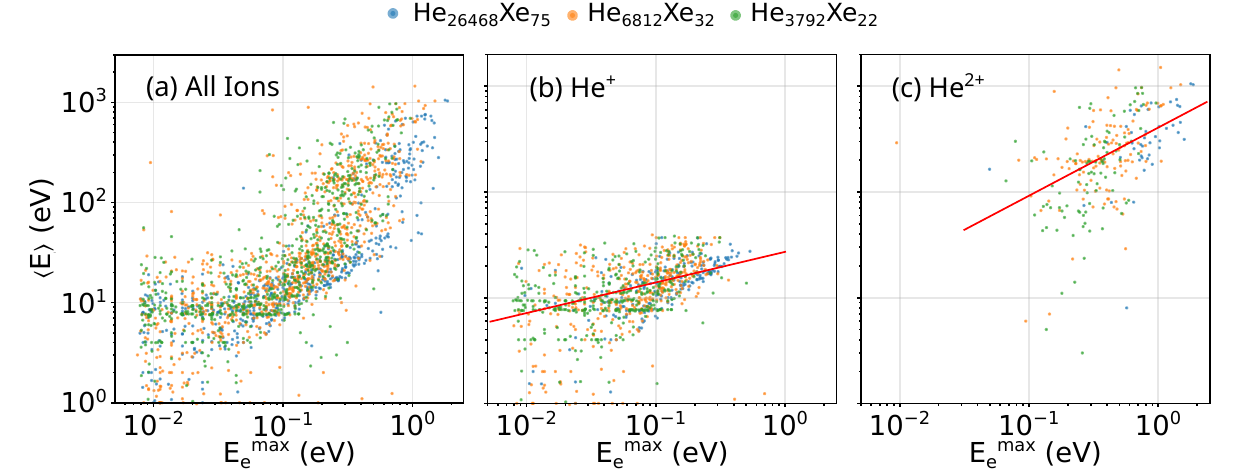}
\caption{
    Panel (a) shows $\Ei$ versus $\Ee$ for individual nanoplasma events (colored dots).
    Blue, orange, and green denote the He$_{26468}$Xe$_{75}$, He$_{6812}$Xe$_{32}$,
    and He$_{3792}$Xe$_{22}$ measurement sets, respectively.
    Panels (b) and (c) select \HeIon-dominated ($W_{\HeIon}\geq0.9$)
    and \HeDIon-dominated ($W_{\HeDIon}\geq0.9$) events, respectively.
    Red lines show power-law fits, $\Ei\propto(\Ee)^n$, to the selected populations.
    \label{fig:ion_ele_ke}
}
\end{figure}

Figure~\ref{fig:ion_ele_ke} shows the average ion kinetic energy $\Ei$ versus
the maximum electron kinetic energy $\Ee$ for individual nanoplasma events.
Panels (b) and (c) select events with $W_{\HeIon}\geq0.9$ and $W_{\HeDIon}\geq0.9$, respectively.
The correlations are fitted with a power law.
For \HeIon, the fit gives $\Ei \propto (\Ee)^{0.3}$.
For \HeDIon, the fit gives $\Ei \propto (\Ee)^{0.6}$.
The correlation suggests that much of the energy gained by electrons during scattering
is transferred to ions via electron-impact ionization.
After losing kinetic energy, electrons recombine with the parent ions.
Although electrons are mostly emitted from highly excited Rydberg atoms, ions
gain kinetic energy through their mutual Coulomb interaction.

\section{Conclusion}

Simultaneous electron VMI and ion TOF measurements allow us to correlate the
ion composition, ion energies, and electron kinetic energies of individual nanoplasma events.
The results reveal a U-shaped distribution in the average charge of emitted ions,
indicating that nanoplasma events predominantly produce \HeIon or \HeDIon ions.
We propose that this U-shaped distribution arises as
a consequence of the Gaussian beam intensity profile in combination with the
ionization dynamics of the nanoplasma.
The MD simulations reproduce this distribution qualitatively, although the relative
abundances differ from those observed experimentally.
Classification by the dominant ion species reveals distinct electron and ion
correlations within the two populations.
Low-energy electron emission is observed in coincidence with \HeIon ions,
whereas the maximum electron kinetic energy is shifted to higher values for events
detected in coincidence with \HeDIon ions. The low-energy electrons are attributed
to the ionization of Rydberg states and the evaporation of electrons from
the Coulomb potential of the cluster.
In contrast, the origin of the shifted maximum kinetic energy in coincidence with
\HeDIon emission remains unclear.
Although it may be related to correlated decay processes in a dense nanoplasma environment,
further investigation is required to understand this behavior fully.
Furthermore, single-shot, single-cluster measurements can be incorporated into existing data pipelines for on-the-fly, online,
or near-online data analysis
\cite{thayerDataSystemsLinac2017, maroneOntheflyDataPostprocessing2017}.
Correlating the detected ion composition with electron and ion signals could provide
an additional diagnostic for beam alignment and event sorting in single-particle
coherent diffractive imaging experiments.
It could also extend nanoplasma-based pump--probe measurements with ion-species information
\cite{medinaLonglastingXUVActivation2023, sishodiaXUVFluorescenceProbe2026}.

\printbibliography

\end{document}


\title{Supplementary Material \\
Correlated detection of electrons and ions emitted from laser-induced nanoplasmas in doped helium nanodroplets in the single-shot single-cluster regime
}
\author{Keshav Sishodia~et.~al.}
\date{}
\maketitle

\section{Laser intensity}
\begin{figure}[ht]
\centering
\includegraphics[width=0.8\linewidth]{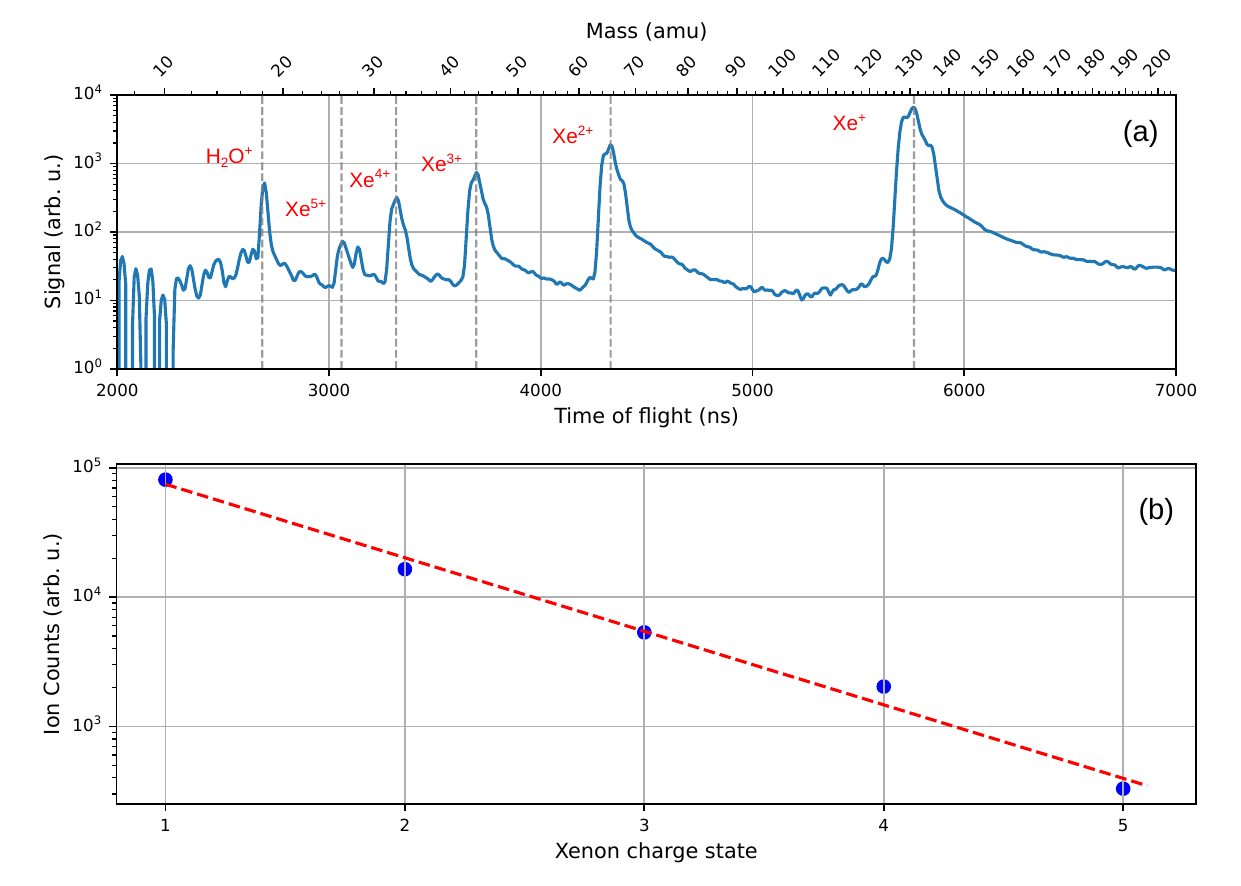}
\caption{
    (a) Ion time-of-flight spectrum of effusive xenon.
    The positions of H$_2$O$^+$, Xe$^{5+}$, Xe$^{4+}$, Xe$^{3+}$, Xe$^{2+}$, and Xe$^{+}$ are marked on the spectrum.
    (b) Total ion counts as a function of xenon charge state.
    The experimental data are fitted with an exponential curve.
    \label{sfig:xenon_spectrum}
}
\end{figure}

To estimate the laser intensity at the focal spot, we recorded the mass spectrum of xenon gas ionized by the strong field of a near-infrared (NIR) beam.
Figure~\ref{sfig:xenon_spectrum} shows the xenon mass spectrum, with charge states up to Xe$^{5+}$.
Based on the experimental curve reported by Augst~\textit{et al.}~\cite{augstLaserIonizationNoble1991},
we estimate the peak NIR intensity to be $1\times10^{15}$~W\,cm$^{-2}$.

\section{Ion time-of-flight spectrometer transmission}

\begin{figure}[ht]
\centering
\includegraphics[width=0.8\linewidth]{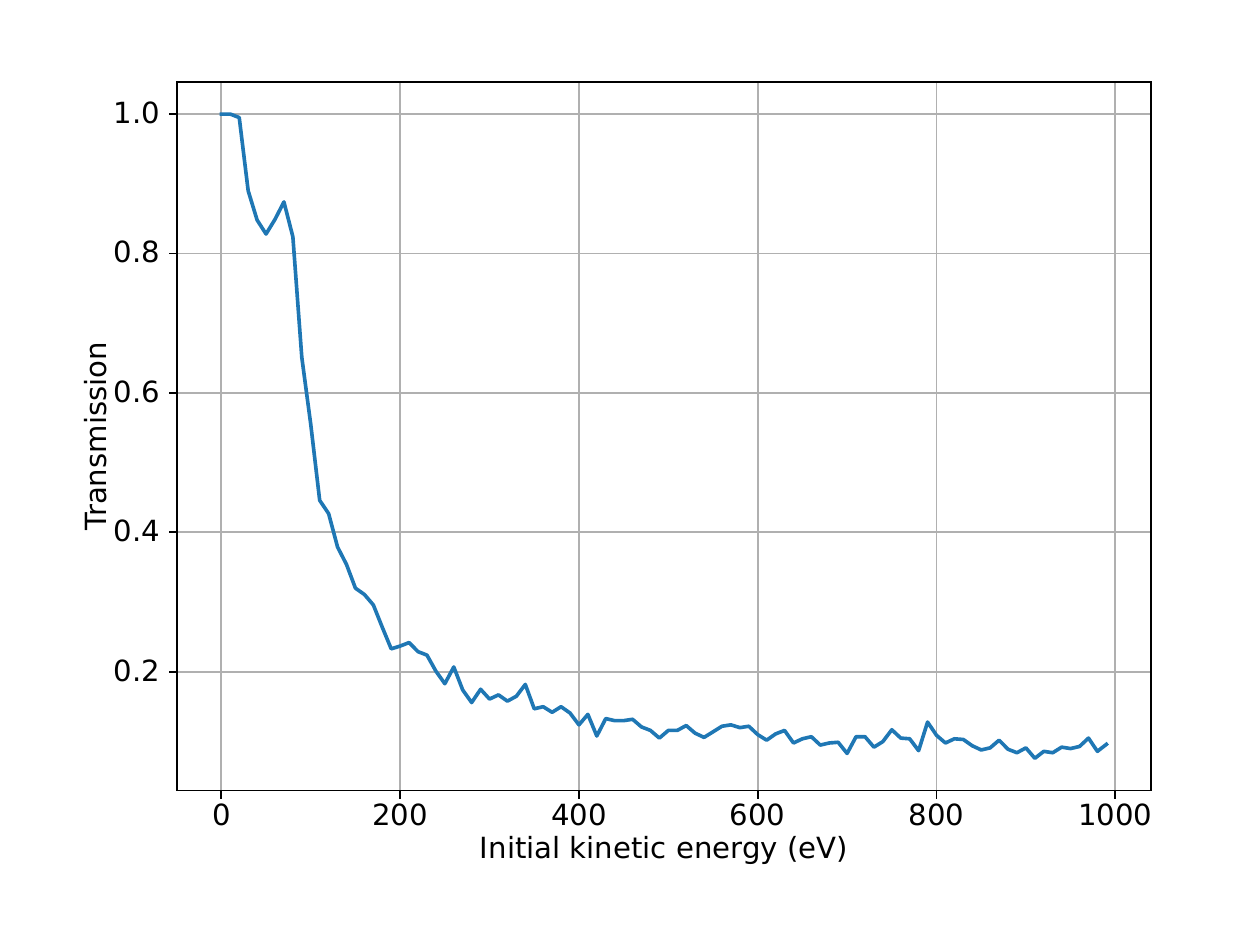}
\caption{
    Transmission of He$^{2+}$ ions through the ion TOF spectrometer.
    For ions with kinetic energies greater than 10~eV, transmission drops significantly.
    \label{sfig:det_trans}
}
\end{figure}

To characterize the transmission of high-energy ions, we simulated their trajectories through the spectrometer.
Figure~\ref{sfig:det_trans} shows the transmission of He$^{2+}$ ions.
The ion TOF spectrometer has 100\% transmission only for ions with low kinetic energies ($<10$~eV).
The transmission gradually decreases to approximately 10\% for ion kinetic energies above 250~eV.

\section{Analysis of individual nanoplasma events}
\label{ssec:event_analysis}

\begin{figure}[ht]
\centering
\includegraphics[width=0.8\linewidth]{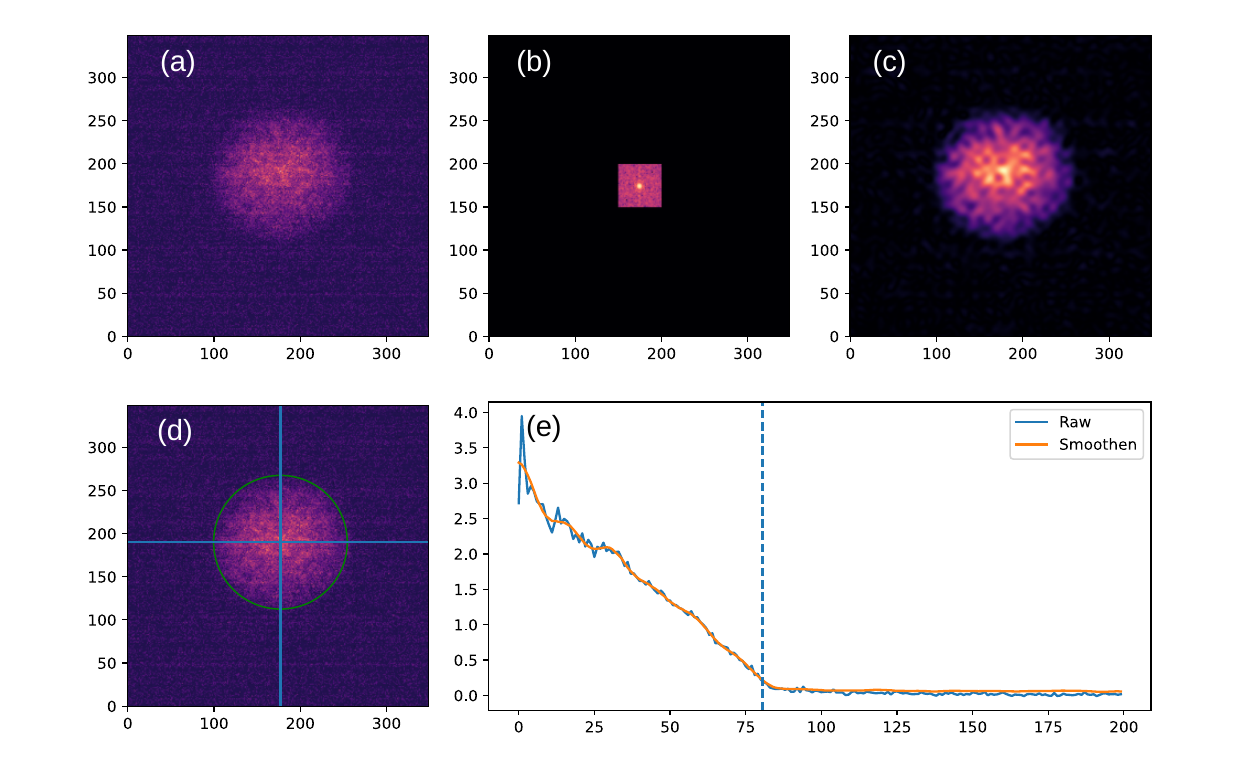}
\caption{
    Single-shot analysis of an electron VMI.
    Panel (a) shows the raw VMI.
    Panel (b) shows the Fourier-transformed image after the removal of high-frequency components to reduce noise.
    Panel (c) shows the filtered image after the inverse Fourier transform.
    Panel (d) indicates the image center and maximum radius, and panel (e) shows the radially integrated signals and the maximum radius.
    \label{sfig:electron_analysis}
}
\end{figure}

Individual nanoplasma events were measured using ion time-of-flight (TOF) and electron velocity-map imaging (VMI) spectrometers, as described in the experimental section of the main text.
Figure~\ref{sfig:electron_analysis} illustrates the analysis of a representative electron VMI.
First, a bright raw VMI was selected based on its signal intensity, as shown in Fig.~\ref{sfig:electron_analysis}(a).
The image was then Fourier-transformed into frequency space, and a low-pass filter was applied to remove high-frequency components.
Figure~\ref{sfig:electron_analysis}(b) shows the filtered image in frequency space, and panel (c) shows the smoothed VMI obtained by applying the inverse Fourier transform.
The smoothed image was used to locate the VMI center and determine the maximum radius ($R_{\mathrm{max}}$) of the circular feature.
Each pixel in the VMI corresponds to an area of $67~\mu\mathrm{m}\times67~\mu\mathrm{m}$.
Images with $R_{\mathrm{max}}\geq5$~px were classified as nanoplasma events.
The VMI center was determined by calculating the autocorrelation of the smoothed image and selecting the brightest pixel.
The standard deviation of the VMI center positions across 2334 images was 1.5~px.
Figure~\ref{sfig:electron_analysis}(d) indicates the center and radius of the circular feature, and panel (e) shows the radial integration of the raw and smoothed VMIs, with a vertical line marking $R_{\mathrm{max}}$.
The radius $R_{\mathrm{max}}$ was then converted to the maximum electron kinetic energy ($E_{\mathrm{e}}^{\mathrm{max}}$).
An error of 1~px in determining $R_{\mathrm{max}}$ would result in an error of 1~meV in $E_{\mathrm{e}}^{\mathrm{max}}$.
The image intensity within the circular feature was integrated to obtain the total electron counts ($N_{\mathrm{e}}$) for each nanoplasma event.

\begin{figure}[ht]
\centering
\includegraphics[width=0.8\linewidth]{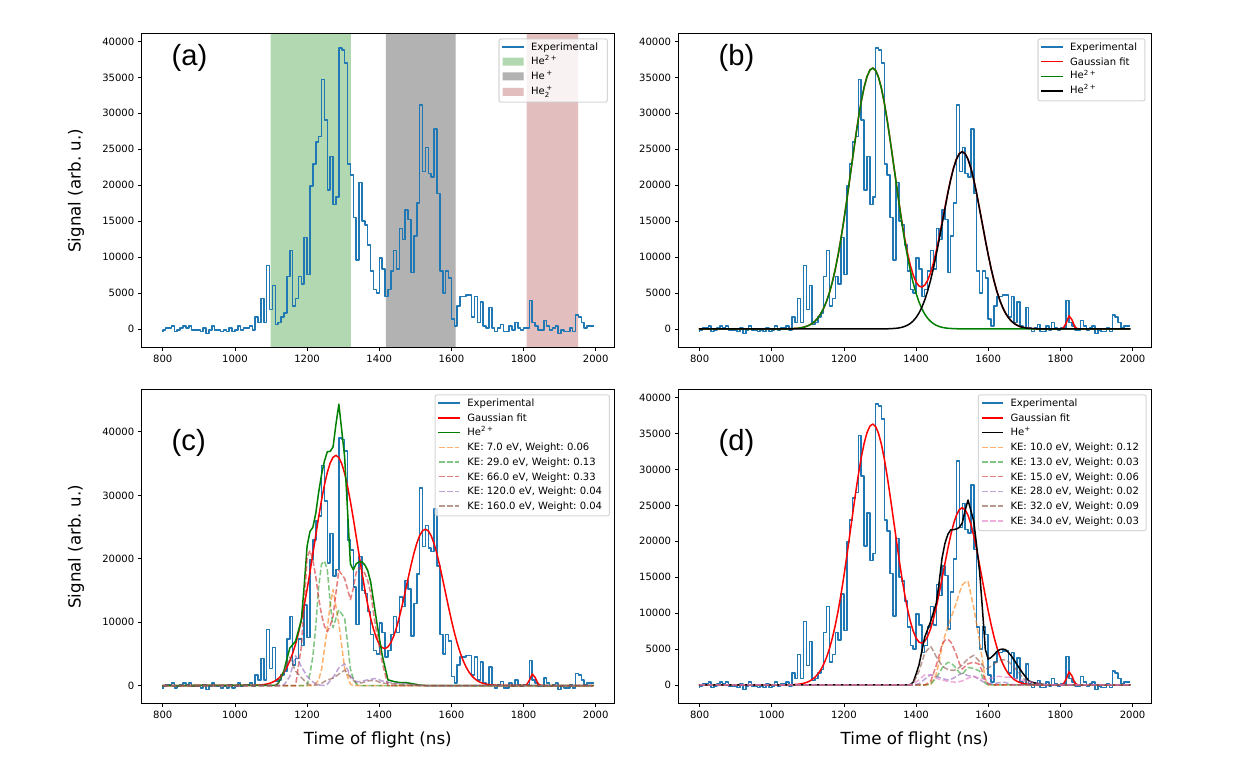}
\caption{
    Single-shot analysis of the ion TOF spectrum.
    Panel (a) shows the ion TOF spectrum, with the expected times of flight for He$^{2+}$, He$^{+}$, and He$_2^+$ ions indicated.
    Panel (b) shows the experimental TOF spectrum fitted with Gaussians.
    Panels (c) and (d) show fits to the Gaussian peaks for He$^{+}$ and He$^{2+}$, respectively, using simulated basis functions.
    Each basis function is shown together with its kinetic energy and fitting weight.
    \label{sfig:ion_analysis}
}
\end{figure}

The single-shot ion TOF spectra were analyzed using a two-step method.
First, each TOF spectrum was fitted with multiple Gaussians to describe the He$^{2+}$, He$^{+}$, and He$_2^+$ ion peaks, as shown in Fig.~\ref{sfig:ion_analysis}(b).
Although this fit gives the relative contributions of the different ion species, it does not directly provide their kinetic energies.
In the second step, a set of basis functions simulated using SIMION~\cite{dahlSimionPersonalComputer2000} was used to fit the Gaussian peaks and determine the kinetic energies of the emitted ions.
Panels (c) and (d) of Fig.~\ref{sfig:ion_analysis} show these fits for He$^{+}$ and He$^{2+}$, respectively, together with the weights of the individual basis functions.
Finally, the average ion kinetic energy ($\langle E_{\mathrm{i}}\rangle$) was calculated as the weighted average of the kinetic energies associated with the basis functions.

\section{Molecular dynamics simulations}

The molecular dynamics (MD) simulation method has been described previously~\cite{heidenreichIonEnergeticsElectronrich2012, heidenreichExtremeIonizationXe2007, heidenreichEfficiencyDopantinducedIgnition2016}.
All atoms and nanoplasma electrons are propagated classically, starting from a cluster of neutral atoms.
Electrons are introduced into the MD simulation when the criteria for tunnel ionization (TI), classical barrier-suppression ionization (BSI), or electron-impact ionization (EII) are met.
These criteria are checked for each atom at every MD time step.
Instantaneous TI probabilities are calculated using the Ammosov--Delone--Krainov formula~\cite{ammosovTunnelIonizationComplex1986},
and EII cross sections are calculated using the Lotz formula~\cite{lotzEmpiricalFormulaElectronimpact1967}, with the ionization energy evaluated relative to the atomic Coulomb barrier in the cluster~\cite{fennelHighlyChargedIons2007}.
Ion-ion interactions are described by Coulomb potentials, whereas electron-ion and electron-electron interactions are described by smoothed Coulomb potentials.
Interactions involving neutral atoms are neglected, except for a Pauli repulsive potential between electrons and neutral He atoms.

Three-body ion-electron recombination (TBR) occurs automatically during a trajectory as the nanoplasma expands and cools sufficiently. To account for this process, trajectories were propagated for 2~ps, except at a pulse peak intensity of $4\times10^{13}$~W\,cm$^{-2}$, where they were propagated for 3~ps to account for the slower nanoplasma expansion. Nanoplasma electrons were classified as recombined if they remained within a cutoff radius of 150~\AA{} of a He ion for at least 500~fs.
To obtain a histogram of $\langle Q\rangle$, a set of trajectories with slightly different initial conditions was simulated, with each trajectory corresponding to a single droplet in the experimental cluster beam.

The simulations were carried out for the He$_{2171}$Xe$_{55}$ droplet.
The simulation results were averaged over the three-dimensional Gaussian intensity profile of the NIR laser focus~\cite{heidenreichKineticEnergyDistribution2011}.
MD simulations were carried out for selected NIR peak intensities $I_M$, using a photon energy of 1.55~eV and a Gaussian temporal intensity envelope with a full width at half maximum (FWHM) of 130~fs.
The sampled intensities ranged from $3\times10^{13}$~W\,cm$^{-2}$, the lowest intensity at which TI of Xe occurred, to $5\times10^{14}$~W\,cm$^{-2}$, at which $\langle Q\rangle$ approached its maximum value of 2.
The range $5\times10^{14}~\mathrm{W\,cm^{-2}}\leq I_M\leq1\times10^{15}~\mathrm{W\,cm^{-2}}$ at the center of the experimental focus was represented by the sample at $5\times10^{14}$~W\,cm$^{-2}$.
For each sampled intensity $I_M>5\times10^{13}~\mathrm{W\,cm^{-2}}$, 100 trajectories were calculated; for each sampled intensity $I_M<5\times10^{13}~\mathrm{W\,cm^{-2}}$, 300 trajectories were calculated.

\printbibliography